\documentclass[%
superscriptaddress,
 amsmath,amssymb,
 prd,
 twocolumn
]{revtex4-2}

\usepackage{graphicx}
\usepackage{dcolumn}
\usepackage{bm}
\usepackage{natbib}
\usepackage{mathrsfs}
\usepackage[normalem]{ulem} 
\usepackage{hyperref}

\usepackage{hyperref}
\hypersetup{
  colorlinks = true,
  urlcolor = blue,
  linkcolor = blue,
  citecolor = green,
  filecolor = magenta,
}

\newcommand{\ud}{\mathrm{d}}
\newcommand{\pd}{\partial}

\newcommand{\wh}{\mathrm{wh}}

\allowdisplaybreaks

\begin{document}

\title{Searching for wormholes with gravitational wave scattering}
\thanks{All authors have contributed equally.}%

\author{Hong Zhang}
\email{hong.zhang@sdu.edu.cn}
\affiliation{
Institute of Frontier and Interdisciplinary Science,
Key Laboratory of Particle Physics and Particle Irradiation (MOE),
Shandong University, Qingdao, Shandong 266237, China}
\author{Shaoqi Hou}
\email{hou.shaoqi@whu.edu.cn}
\affiliation{School of Physics and Technology, Wuhan University, Wuhan, Hubei 430072, China}
\author{Shou-shan Bao}
\email{ssbao@sdu.edu.cn}
\affiliation{
Institute of Frontier and Interdisciplinary Science,
Key Laboratory of Particle Physics and Particle Irradiation (MOE),
Shandong University, Qingdao, Shandong 266237, China}

\date{\today}

\begin{abstract}

We propose using the gravitational wave scattering off spherical wormholes to search for their existence. We carefully calculate the reflected and transmitted waveforms with time-independent scattering theory. Our results quantitatively show the echo signatures in the two universes on both sides of the wormhole. In a certain wormhole mass range, the transmitted wave has a unique isolated chirp without an inspiral waveform, and the reflected wave has the anti-chirp behavior, i.e., the missing of the chirping signal. We also calculate the searching range of the current and projected gravitational wave telescopes.  Our method can be adapted to efficiently calculate the templates to search for wormholes.

\end{abstract}

\maketitle


\section{Introduction}

Since its birth, general relativity has always brought human beings with fascinating ideas and astonishing phenomena. Among them, wormholes are probably one of the most bizarre objects. They are ``bridges'' or ``handles'' connecting regions of our universe at distances, or even two universes. They were firstly conjectured by Einstein and Rosen, originally to model elementary particles \cite{Einstein:1935tc}. After that, Misner and Wheeler gave them the name that we call them today \cite{Misner:1957mt}. In 1973, Ellis \cite{Ellis:1973yv} and Bronnikov \cite{Bronnikov:1973fh} independently found traversable wormhole solutions. Later, Morris and Thorne \cite{Morris:1988cz,Morris:1988tu} and Visser \cite{Visser:1989kh}  aroused general interest in these solutions in the literature. Although being at the center of attention for so many years, wormholes have not been spotted yet.

As another important prediction of general relativity, the gravitational wave (GW) has been confirmed by the observations made by LIGO-Virgo-KAGRA collaborations \cite{Abbott:2016blz,TheLIGOScientific:2017qsa,LIGOScientific:2021qlt}. Up to now, there have been 90 GW events and more alerts detected \cite{LIGOScientific:2018mvr,LIGOScientific:2020ibl,LIGOScientific:2021usb,LIGOScientific:2021djp}. The GW is not merely a new tool to probe the nature of gravity in the high speed and dynamical regime \cite{Berti:2015itd,Gong:2018cgj,Hou:2020tnd}, but offers the opportunity to search for exotic astrophysical objects, such as firewall, fuzz balls, boson stars, gravastars, and of course, wormholes. These exotic compact objects mimic black holes in the sense that their radii are close to the Schwarzschild radii of black holes with the same masses. But they produce additional echoes after the ringdown phase of the binary system evolution \cite{Mark:2017dnq}. Wormholes are special because they possess an effective double-peak potential. The GW bounces back and forth between the two peaks while leaking out of the wormhole, producing a series of echoes \cite{Cardoso:2016rao,Cardoso:2016oxy,Bueno:2017hyj,Ghersi:2019trn}. Other methods of searching for wormholes include detecting the anti-chirp signal produced by a small black hole crossing the wormhole throat \cite{Dent:2020nfa},
observing the echoes of electromagnetic signals traversing wormholes \cite{Liu:2020qia}, and measuring the anomalous motion of objects on one side of the throat affected by charges and masses on the other side \cite{Dai:2019mse,Simonetti:2020vhw}.

The scattering of external GWs off a wormhole does not depend on the matter distribution in the wormhole, especially close to the throat, making it a model-independent method to search for wormholes. It has been studied with a Gaussian wave packet and complicated numerical programs solving the time evolution \cite{Ghersi:2019trn}. In this work, we apply the time-independent scattering theory which greatly reduces the complexity of the calculation as well as the numerical error. The calculation is thus much faster and more straightforward, enabling us to use real-life input GWs generated by \verb+PyCBC+ \cite{alex_nitz_2019_2643618s}. In the rest of this article, we first carefully investigate the features of the reflected and transmitted GW signals.  Then we quantitatively study the probability of searching for wormholes with masses from $10M_\odot$ to $10^3 M_\odot$ by calculating the signal-to-noise ratio (SNR). Our method can be adapted to efficiently calculate the templates for GW telescopes.

This paper is organized as follows. In Sec.~\ref{sec:framework}, we present the perturbation theory of wormhole metric and our calculation framework. In Sec.~\ref{sec:results}, we show the features in the calculated reflected and transmitted signals, including the isolated-chirp, anti-chirp and the echoes. The searching scopes of wormholes by the current and projected GW telescopes are also presented. A short summary is given in Sec.~\ref{sec:summary}.

\section{Wormhole Perturbation Theory}\label{sec:framework}

\subsection{Time-independent framework}

We consider a simple kind of wormhole by sewing two identical Schwarzschild metrics at $r_0=r_g+\Delta r_0>r_g=2 GM$. Each Schwarzschild metric is described by
\begin{equation}
  \label{eq-met}
  \ud s^2=- f(r) \ud t^2+f^{-1}(r) \ud r^2 +r^2(\ud\theta^2+\sin^2\theta \ud\phi^2),
\end{equation}
where $f(r)=1-r_g/r$. The discontinuity at the throat requires the presence of a thin shell of matter \cite{Visser:1989kh}. This metric is asymptotically flat at infinity.

 \begin{figure}
    \includegraphics[clip, trim=0cm 3cm 0cm 4cm, width=0.4\textwidth]{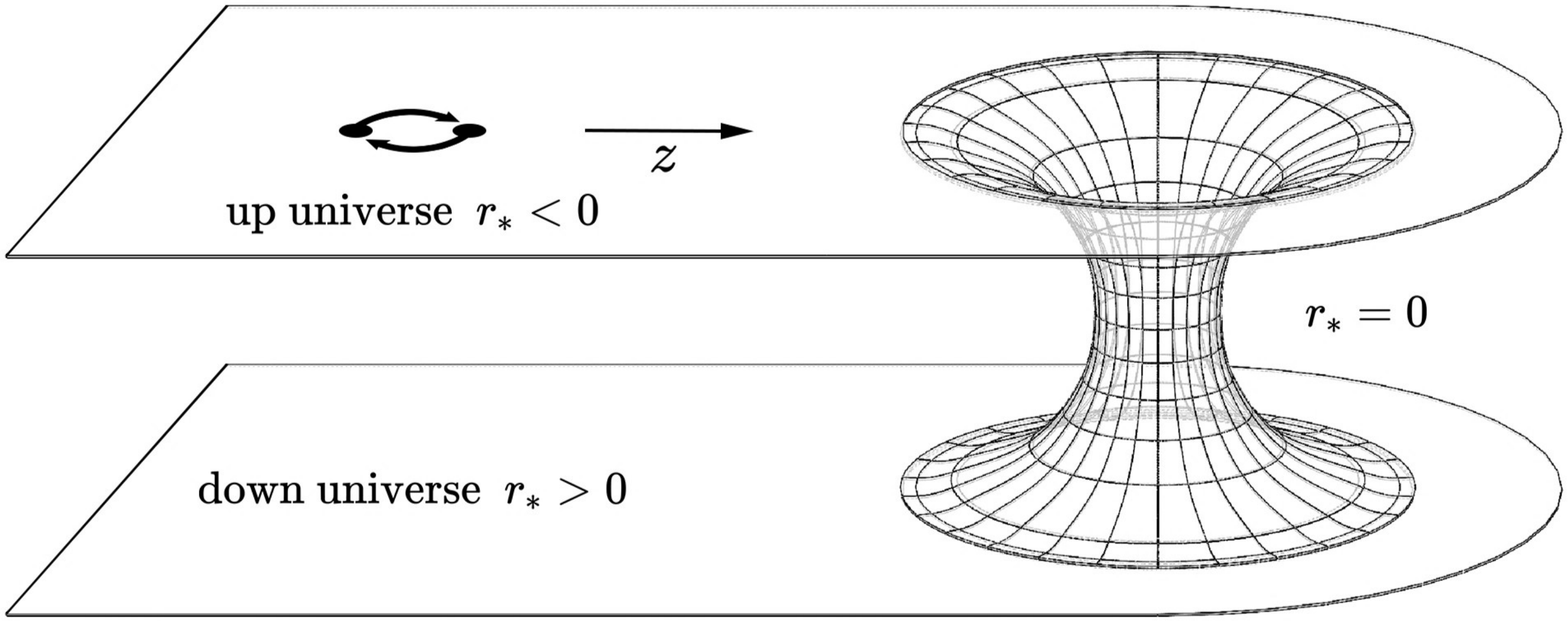}
    \caption{A schematic plot of external GW scattering off a wormhole.
    \label{fig:schematic}}
\end{figure}

We refer to the region in which the GW source resides as the up universe, and call the other side the down universe. The wormhole connects these two regions. In the down universe, one could define the tortoise radius,
\begin{equation}
  \label{eq-def-rs}
  r_*=r-r_0+r_g\ln\frac{r-r_g}{r_0-r_g},
\end{equation}
which is zero at the throat and increases monotonically with $r$. The radius $r_*$  could be analytically continued to the up universe by adding an overall minus sign to the right-hand side of Eq.~\eqref{eq-def-rs}. With this definition, the $r_*$ is negative in the up universe where the GW source resides, positive in the down universe, and zero only at the throat of the wormhole. A schematic plot is shown in Fig.~\ref{fig:schematic}.

\begin{figure}
    \centering
    \includegraphics[clip, width=0.4\textwidth]{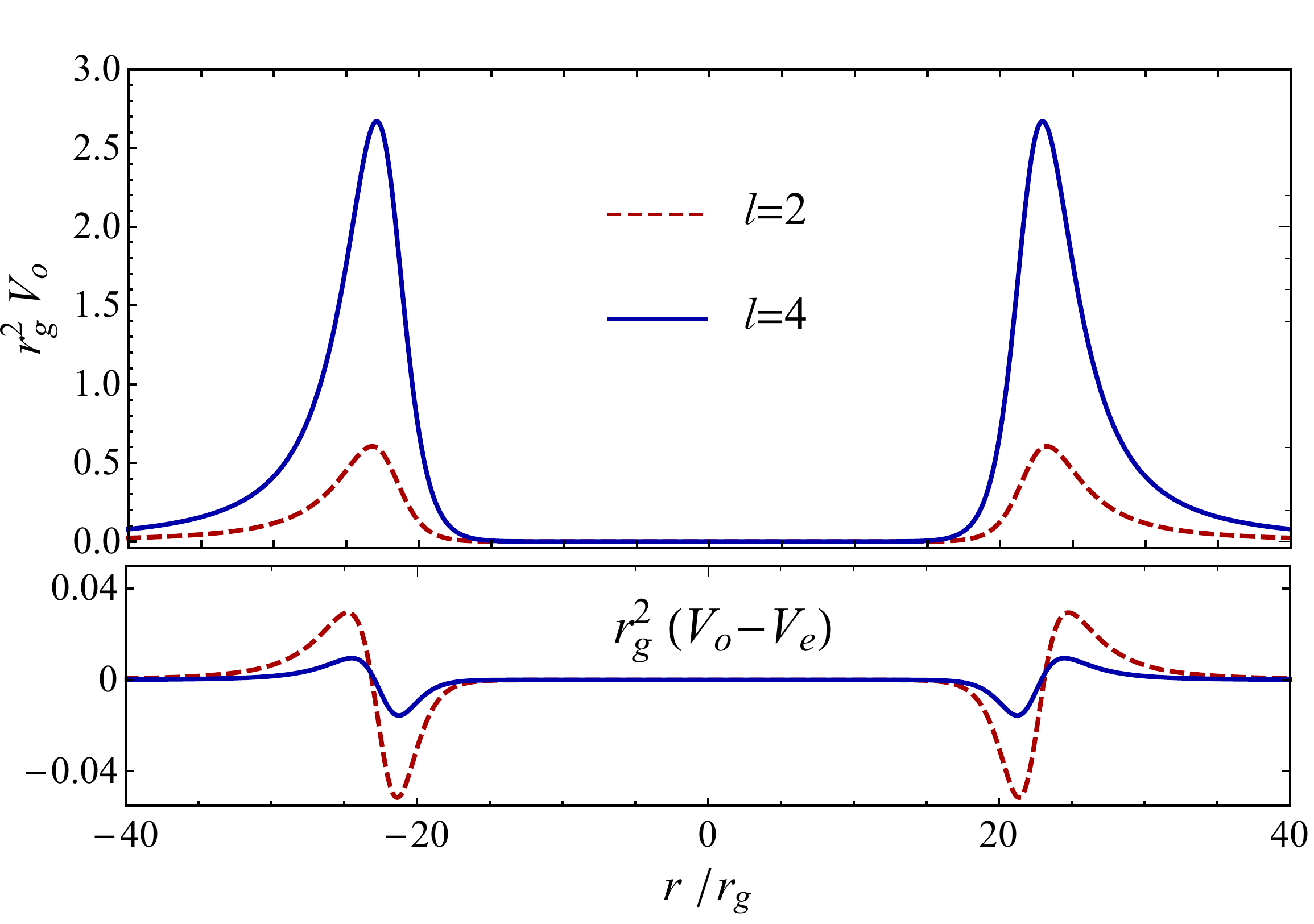}
    \caption{The odd parity potential $V^l_\text{o}$ (upper) and the difference $V^l_\text{o}-V^l_\text{e}$ (lower) normalized by $1/r_g^2$, with $l=2$ and $4$, respectively.  
    \label{fig:pot}}
\end{figure}

The gravitational perturbation close to a spherical wormhole can be studied by generalizing the mathematical framework for Schwarzschild black holes  \cite{Regge:1957td,Zerilli:1970wzz,Chandrasekhar:1985kt,Martel:2005ir}. The perturbation $h_{\mu\nu}$ is firstly expanded with spherical partial waves specified by $l$, $m$, and parity, which can be found in Appendix.~\ref{app:plane-wave}. The gauge-dependent expansion coefficients can then be combined linearly to construct gauge-invariant even-parity Zerilli-Moncrief function $\Psi_\text{e}^{lm}$ and odd-parity Regge-Wheeler function $\Psi_\text{o}^{lm}$. Their time evolutions are described by \cite{Martel:2005ir},
\begin{equation}
  \label{eq-mseq-e}
  \left(-\pd_t^2+\pd_{r_*}^2-V^{l}_\text{e/o}\right)\Psi_\text{e/o}^{lm}=0,
\end{equation}
where the subscript stands for the even or odd parity, and the partial-wave potentials are,
\begin{subequations}
\begin{align}
V_\text{o}^{l} &= f(r) \left[ \frac{l(l+1)}{r^2}-\frac{3r_g}{r^3} \right],  \label{eq-def-vo}\\
V^{l}_\text{e} &= \frac{f(r)}{\Lambda^2}\left[ \mu^2 \left( \frac{\mu+2}{r^2}+\frac{3r_g}{r^3} \right) +\frac{9r_g^2}{r^4} \left( \mu+\frac{r_g}{r} \right) \right], \label{eq-def-ve}
\end{align}
\end{subequations}
with $\mu=(l-1)(l+2)$ and $\Lambda=\mu+{3 r_g}/{r}$. 
In Fig.~\ref{fig:pot}, we show the double-peak structures of the potentials. The value of $\Delta r_0$ is chosen as $10^{-8}r_g$. The heights of the peaks in the potentials increase with $l$ for both odd and even parities. For large $l$, it depends approximately on $l^2$ . The peaks of even parity potential are a little higher than those of odd parity with the same $l$. The percentage difference of $V^l_\text{o}$ and $V^l_\text{e}$ is of order $10^{-5}$ for $l=2$ and decreases as $l^{-3}$ when $l$ increases. The difference barely matters for $l>10$. 

The contribution of the matter at the wormhole throat is ignored in Eq.~\eqref{eq-mseq-e}. In this work, we consider an GW wave packet from outside of the wormhole. The incident GW perturbs both the wormhole metric and the matter supporting its throat. Their feedback to the incident GW is at the second order of $h_{\mu\nu}$. Therefore, the master equation in Eq.~\eqref{eq-mseq-e} is sufficient for our purpose \cite{Cardoso:2016rao,Cardoso:2016oxy,Bueno:2017hyj,Dent:2020nfa}.

The problem is now reduced to a one-dimensional scattering problem, which is more convenient to handle in the frequency space,
\begin{equation}
  \label{eq-def-four}
  \Psi_\text{e/o}^{lm}(t,r_*)=\int\frac{\ud\omega}{2\pi}\widetilde\Psi_\text{e/o}^{lm}(\omega,r_*)e^{-i\omega t}.
\end{equation}
We normalize the eigenstate $\widetilde{\Psi}_\text{e/o}^{lm}$ with frequency $\omega$ as,
\begin{equation}\label{eq:Psi-eigenstate}
   \widetilde\Psi^{lm}_\text{e/o}\rightarrow \left\{ 
	\begin{array}{cl}
	{e^{i \omega r_*}} + \mathcal{R}^{l}_\text{e/o} e^{-i\omega r_*}, & \quad r_*\rightarrow -\infty,\\
	\mathcal T^{l}_\text{e/o} e^{i\omega r_*}, &\quad r_*\rightarrow+\infty.
    \end{array}\right.
\end{equation}
Both the Wentzel–Kramers–Brillouin (WKB) approximation and the numerical Wronskian method are used in the calculation of $\mathcal{R}$ and $\mathcal{T}$  (see Appendix~\ref{app:WKB} for details).
With all eigenstates at hand, the initial wave packet could be written as a linear combination of these eigenstates at some early time.  Then the later evolution of the wave packet is fully controlled by the time-evolution of the eigenstates.

\subsection{Initial Wave Packet}

We still need the initial conditions for $\Psi_\text{e/o}^{lm}$ at some early time $t_i$. Since the wormhole metric is asymptotically flat and the source is far away from the wormhole, the initial GW can be considered as a packet of plane waves. We calculate the $\widetilde{\Psi}_\text{e/o}^{lm}$ for each plane wave with frequency $\omega$. Then the initial values $\Psi_\text{e/o}^{lm}$ of the packet are expressed as integrals of these scattering eigenstates. With the time evolution of each eigenstate, the time evolutions of $\Psi_\text{e/o}^{lm}$ are straightforward. Finally, the two observed polarizations $h^{\mathcal{R/T}}_+$ and $h^{\mathcal{R/T}}_\times$ are calculated after the wave packet scatters off the wormhole and propagates to infinity.

At some early time $t_i$, the incident GW packet $h_{\mu\nu}(z=-r_* \cos\theta,t_i)$ locates in a finite range close to some $z_i$. Since the wormhole metric is asymptotically flat, the wave packet can be expanded in terms of plane waves at this time,
\begin{align}\label{eq:in-wave-PW-expansion}
h_{\mu\nu}(t_i) 
&=\sum_{P=\pm} \int_{-\infty}^\infty \frac{\ud\omega}{2\pi} \, \widetilde{h}_P(\omega) e^{(P)}_{\mu\nu}  e^{i\omega (z-t_i)},
\end{align}
where the dependence on $r_*$ and $\theta$ on the left-hand side is suppressed for compactness, the transverse-traceless gauge is adopted implicitely, $\widetilde{h}_P(\omega)$ is the amplitude in the frequency space, and $e^{(P)}_{\mu\nu}$ are the helicity basis, with $P=-$ and $+$ representing the left- and right-handed polarizations respectively. Since $h_{\mu\nu}$ is real, one has $\widetilde{h}_P(\omega) = \widetilde{h}^*_{-P}(-\omega)$. We use the method described in Ref.~\cite{Martel:2005ir} (also see the lecture note in Ref.~\cite{Berti:note}) to match the plane wave $e^{(P)}_{\mu\nu}e^{-i\omega r_* \cos\theta}$ to the gauge invariant functions $\widetilde\Psi_{\text{e/o}}^{lm}$, which is explained in detail in Appendix~\ref{app:plane-wave}.
Then the initial conditions $\Psi^{lm}_{\text{e/o}}(t_i)$ are obtained by integrating  $\widetilde\Psi_{\text{e/o}}^{lm}$ for plane waves with $\widetilde{h}_P(\omega)$.
After some algebra, one obtains the $\Psi_{\text{o}}^{lm}$ at $t_i$,
\begin{align}\label{eq:Psi-o-ti}
\Psi_\text{o}^{lm}(t_i)
&= 
\int_{-\infty}^{+\infty} \frac{\ud\omega}{2\pi}
\mathcal{A}_\text{o}^{lm}(\omega)
e^{i\omega(r_*-t_i)},
\end{align}
where,
\begin{align}\label{eq:A-o}
\mathcal A_\text{o}^{lm}(\omega)=-\frac{i^{l+1}}{\sqrt 2}m_l\sum_{P=\pm}\mathcal F_{lm}^P i^P \frac{\tilde h_P(\omega)}{\omega},
\end{align}
where  $m_l=[(l+2)(l+1)l(l-1)]^{-1/2}$, $\mathcal F_{lm}^P=-\delta_{m,2P}i^{l}\sqrt{2\pi(2l+1)}e^{-2iP\psi}$ and $\psi$ is the polarization angle, which is set to zero. The Kronecker delta sets the value of $m$ to be $\pm 2$, which constrains the value of $l$ to be larger or equal to $2$. This function is nonzero only in the neighborhood of some $r_*< 0$ corresponding to $z_i$. The expression for even parity is almost the same as Eqs.~\eqref{eq:Psi-o-ti} and \eqref{eq:A-o}, only with the $i^P$ factor dropped in Eq.~\eqref{eq:A-o}. These are the initial conditions at time $t_i$. The later time evolution is then straightforward with the eigenstates of Eq.~\eqref{eq-mseq-e}, with time derivative replaced by $-i\omega$. One could simply change $e^{i\omega r_*}$ in Eq.~\eqref{eq:Psi-o-ti} to $\widetilde\Psi^{lm}_\text{o/e}$ normalized as Eq.~\eqref{eq:Psi-eigenstate}, and multiply the integrand by $e^{-i\omega(t_f-t_i)}$ for the time evolution of the eigenstates from $t_i$ to some final time $t_f$. After subtracting the incident plane GW in the up universe, one obtains reflected $\Psi_\text{e/o}^{lm}$ at $t_f$,
\begin{align}\label{eq:Psi-R-tf}
\Psi_{\text{e/o};\mathcal{R}}^{lm}(t_f)
&= 
\!\!\!
\int_{-\infty}^{+\infty}\!\! \frac{\ud\omega}{2\pi}
\mathcal{A}_\text{e/o}^{lm}(\omega)
\!\!
\left[\mathcal{R}^l_\text{e/o} \! - \! (-1)^{l+1}\right]
\!
e^{-i\omega(r_*+t_f)},
\end{align}
with $r_* \ll 0$ and the transmitted wave,
\begin{align}\label{eq:Psi-R-tf}
\Psi_{\text{e/o};\mathcal{T}}^{lm}(t_f)
&= 
\!\!\!
\int_{-\infty}^{+\infty}\!\! \frac{\ud\omega}{2\pi}
\mathcal{A}_\text{e/o}^{lm}(\omega)
\mathcal{T}^l_\text{e/o}
e^{i\omega(r_*-t_f)},
\end{align}
with $r_* \gg 0$. The magnitudes of the two polarizations could be written in terms of these $\Psi$'s, \cite{Martel:2005ir}
\begin{subequations}\label{eq:h}
\begin{align}
h_+^{\mathcal R/\mathcal T} =& \frac{1}{r}
\sum_{lm}  \left\{ \Psi_{\text{e;}\mathcal R/\mathcal T}^{lm}
\left[\frac{\partial^2}{\partial\theta^2} +\frac{1}{2} l(l+1)\right] Y^{lm}\right.\nonumber\\
&\left.-\Psi_{\text{o;}\mathcal R/\mathcal T}^{lm} \frac{im}{\sin\theta}\left[ \frac{\partial}{\partial \theta} -\frac{\cos\theta}{\sin\theta}\right] Y^{lm}\right\},\label{eq:hplus}\\
h_\times^{\mathcal R/\mathcal T} =&\frac{1}{r}
\sum_{lm}  \left\{ \Psi_{\text{e;}\mathcal R/\mathcal T}^{lm}
\frac{im}{\sin\theta} \left[ \frac{\partial}{\partial \theta} -\frac{\cos\theta}{\sin\theta}\right]
 Y^{lm}\right.\nonumber\\
&\left.+\Psi_{\text{o;}\mathcal R/\mathcal T}^{lm} 
\left[\frac{\partial^2}{\partial\theta^2} +\frac{1}{2} l(l+1)\right] Y^{lm}\right\}.\label{eq:htimes}
\end{align}
\end{subequations}
Then one could calculate the responses of GW telescopes to these signals.

 \begin{figure}
    \includegraphics[width=0.4\textwidth]{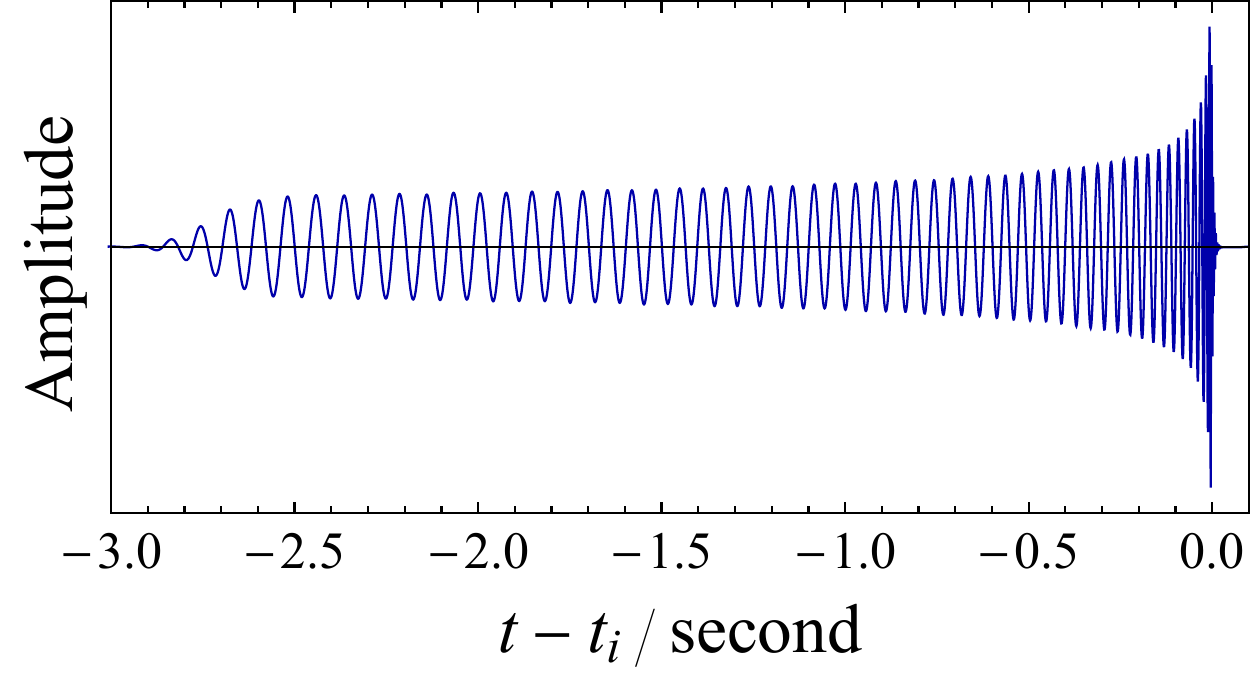}
    \caption{The input incident waveform is generated by \texttt{PyCBC} assuming the spinless black holes have masses the same as GW150914. A cut-off at $-2.6$~second is added by hand, which is checked not affecting the results presented in this paper. The normalization of the amplitude is arbitrary.
    \label{fig:incident}}
\end{figure}

In this work, we generate the incident GW using \verb+PyCBC+ \cite{alex_nitz_2019_2643618s} assuming that the spinless black holes have masses $m_1=35.6~M_\odot$ and $m_2=30.6~M_\odot$. These masses are from the GW150914 \cite{LIGOScientific:2016aoc}, hence the output GW represents a real-life source. The generated waveform is shown in Fig.~\ref{fig:incident}. The considered frequency band of $\omega/2\pi$ is between 5~Hz and 614~Hz, to which LIGO is sensitive. A mild cut-off is added by hand at $-2.6$~second, which is checked not influencing the results presented in this paper. We have calculated the reflected and transmitted waveforms for wormhole masses between $10 M_\odot$ and $10^3 M_\odot$. The parameter $\Delta r_0$ is chosen to be $10^{-8} r_g$. The effect of varying $\Delta r_0$ will be discussed at the end.

\section{Results and Discussions}\label{sec:results}

\begin{figure}
    \centering
    \includegraphics[clip, trim=1cm 0cm 0cm 0cm, width=0.45\textwidth]{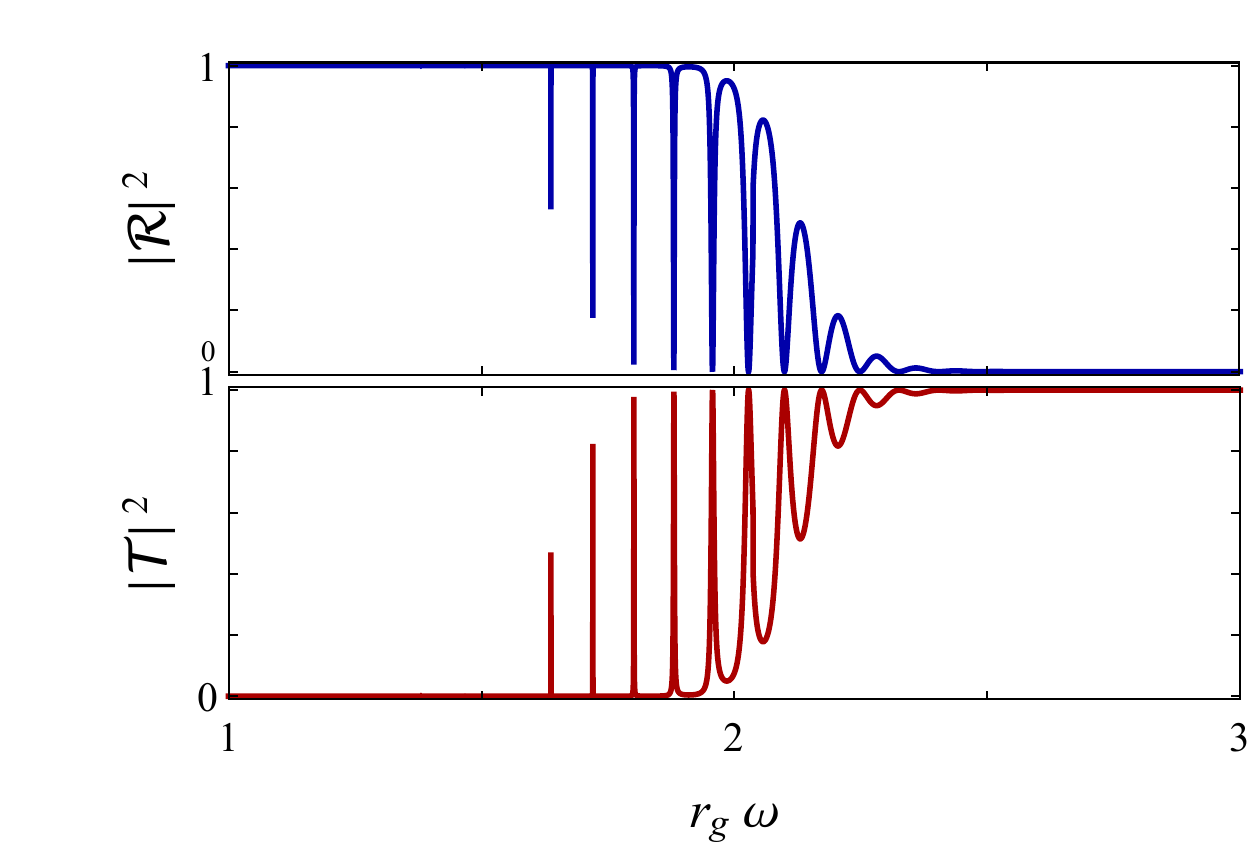}
    \caption{The squared amplitudes of the reflection coefficient (upper) and the transmission coefficient (lower) for $V^l_\text{o}$ with $l=5$, as a function of $r_g \omega $. The critical $ r_g \omega_c$ is $2.037$. For all values of $l<1000$, the resonance region is between $\text{max}\{0,r_g \omega_\text{c}-3\}$ and $r_g\omega_\text{c}+3$, beyond which the incident wave is dominantly reflected or transmitted.
    \label{fig:rftr}}
\end{figure}

We first analyze a single-frequency wave scattering off the double-barrier potential $V_\text{e/o}^l$. Since the detector locates far away from the wormhole, only the coefficients $\mathcal{R}_\text{e/o}^{l}$ and $\mathcal{T}_\text{e/o}^{l}$ defined in Eq.~\eqref{eq:Psi-eigenstate} are relevant to observation. As shown in Fig.~\ref{fig:pot}, the potentials scale as $1/r_g^2$, which is inversely proportional to the cross-section of the corresponding Schwarzschild horizon. Physically, a heavier wormhole has a larger opening and lower potentials, making the transmission relatively easier. The height of the peaks determine a critical frequency $\omega_\text{c}$ which equals the square root of the maximum value of the potential. 
For odd parity potential, the values of $r_g\omega_\text{c}$ are $0.77791, 1.21922$ and $1.63384$ for $l=2,3,4$, respectively. The values for even parity potentials are $0.77797, 1.21922$ and $1.63384$.
If the frequency $\omega$ of the incident wave is much smaller than $\omega_\text{c}$,  the wave scarcely passes through the barriers  to the other side of the wormhole. On the other hand, the incident waves with $\omega\gg \omega_\text{c}$ dominantly transmit. Interesting resonant phenomena happen when $\omega$ is at the same order of $\omega_\text{c}$, a result of the double-peak configuration of the potentials. As an illustration, the squared magnitudes of the reflection and transmission coefficients for $l=5$ are shown in Fig.~\ref{fig:rftr}. The peaks are the discrete frequencies where the incident wave is dominantly reflected or transmitted because of the resonance effect.

For the GW emitted by a binary black hole system,  its frequency increases gradually in the inspiral phase, reaches the maximum at the merger phase, and decreases in the ringdown phase. The amplitude also increases gradually at first, chirps, and then, plunges to zero. For a partial wave with some typical value of $l$, at first the frequency is so small that the wave is nearly completely reflected. When the frequency increases to the resonance region, the wave is partially reflected and partially transmitted. Finally, when the frequency increases to the right of the resonance region as shown in Fig.~\ref{fig:rftr}, transmission dominates. Therefore, for this typical partial wave, one expects that the transmitted wave contains the later part of the inspiral phase and the chirp, while the reflected wave contains the early stage of the inspiral phase and the ringdown phase. Series of echoes also exist after the ringdown phase for both the transmitted and reflected waves. These signatures facilitate the search for the wormholes.

The reflected waves can be intuitively considered as a superposition of the {\it direct} reflection from the left peak of the potential (see Fig.~\ref{fig:pot}) and the {\it secondary} reflection from the right peak with a time delay compared to the direct one. There are more reflections from bouncing back and forth between the two barriers, but they are too small to be dominant. 
At the very early stage, the reflected wave has only the direct reflection. The secondary reflection does not contribute because of the time delay.
Later,\ the direct and secondary reflections overlap and interfere in the reflected wave, resulting a beat-like structure \cite{Hou:2019dcm}. This is observed in our calculation for each partial wave. However, in this region the convergence of the $l$-summations in the calculation of $h^{\mathcal{R}}_+$ and $h^{\mathcal{R}}_\times$ are very slow, and a smarter strategy has to be applied. 
In this work, we focus on the early stage where only the direct reflection contributes. The signal is consequently clean for observation. 
This convergence problem does not exist for transmitted waves.


\begin{figure*}
    \centering
    \includegraphics[width=\textwidth]{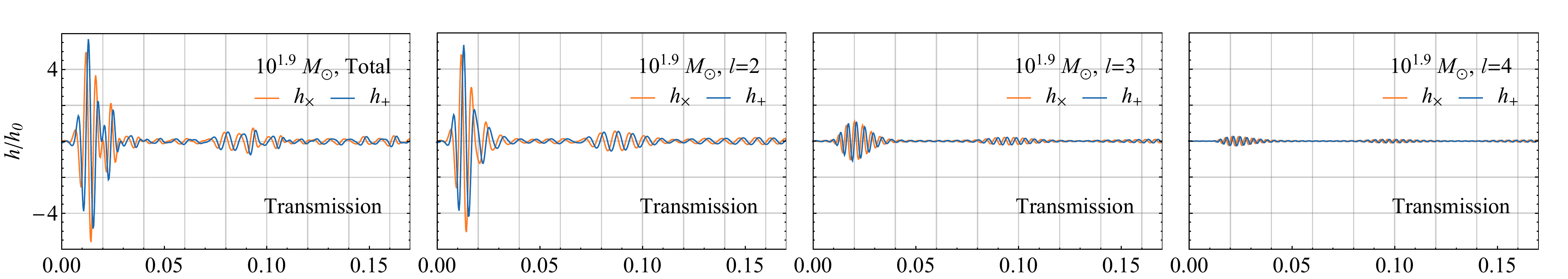}\\
    \includegraphics[width=\textwidth]{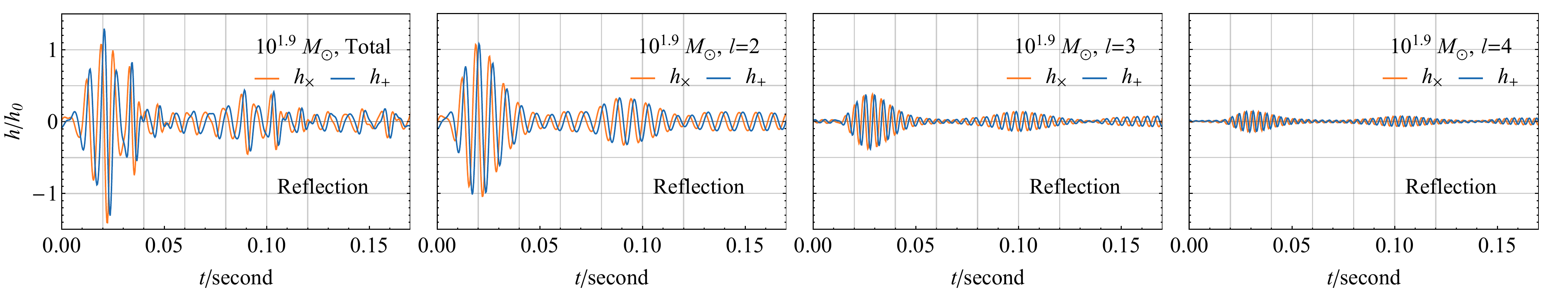}
    \caption{
    The waveforms for the transmitted (upper) and reflected (lower) waves as well as the lowest 3 partial waves. The parameters are $M_{\wh}=10^{1.9}M_\odot$, $\Delta r_0=10^{-8}r_g$, $\theta=\pi/3$ and $\phi=0$. The leftmost plots in both panels are the superposition of partial waves from $l=2$ to $12$. Partial waves with $l>12$ are too small and ignored. The corresponding values of $\omega_c/2\pi$ of $V_\text{o}^l$ are $143.4$~Hz, $224.8$~Hz, and $301.3$~Hz for $l=2,3,4$, respectively. The $t=0$ is reset to the time that the direct and secondary reflected waves from different stages of the incident wave train stop to overlap for an observer (see text). For the reflected waveforms with $t>0$, the $l$-summation converges fast and the results are reliable. This convergence problem does not exist for the transmitted wave. The amplitudes depend on the distance between the source and the wormhole, as well as the distance between the wormhole and the observer. We have divided a common factor $h_0$ for normalization in all the figures.}
    \label{fig:ref_tran_38}
\end{figure*}

Fig.~\ref{fig:ref_tran_38} shows the obtained transmitted and reflected waveforms with the wormhole mass $M_\text{wh} = 10^{1.9}M_\odot$ and $\Delta r_0 = 10^{-8}r_g$. For the transmitted wave, the peaks before $t=0.03$~second are the direct transmission of the chirp. The first echo starts at about $0.07$~second. To our best knowledge, this ``isolated-chirp'' signal without inspiral phase does not exist for any other GW source and can be used as a unique signature of wormholes. For reflected wave, the first echo starts from $t \sim 0$~second. Especially, there is no chirp in the reflected wave. This absence is often called ``anti-chirp". In both panels, the echoes in the reflected and transmitted partial waves are well separated. Nonetheless, the shape of the echoes after summing all partial waves becomes irregular. The first echo is still clear, but the later echoes are not very recognizable.

\begin{figure}
    \centering
    \includegraphics[width=0.232\textwidth]{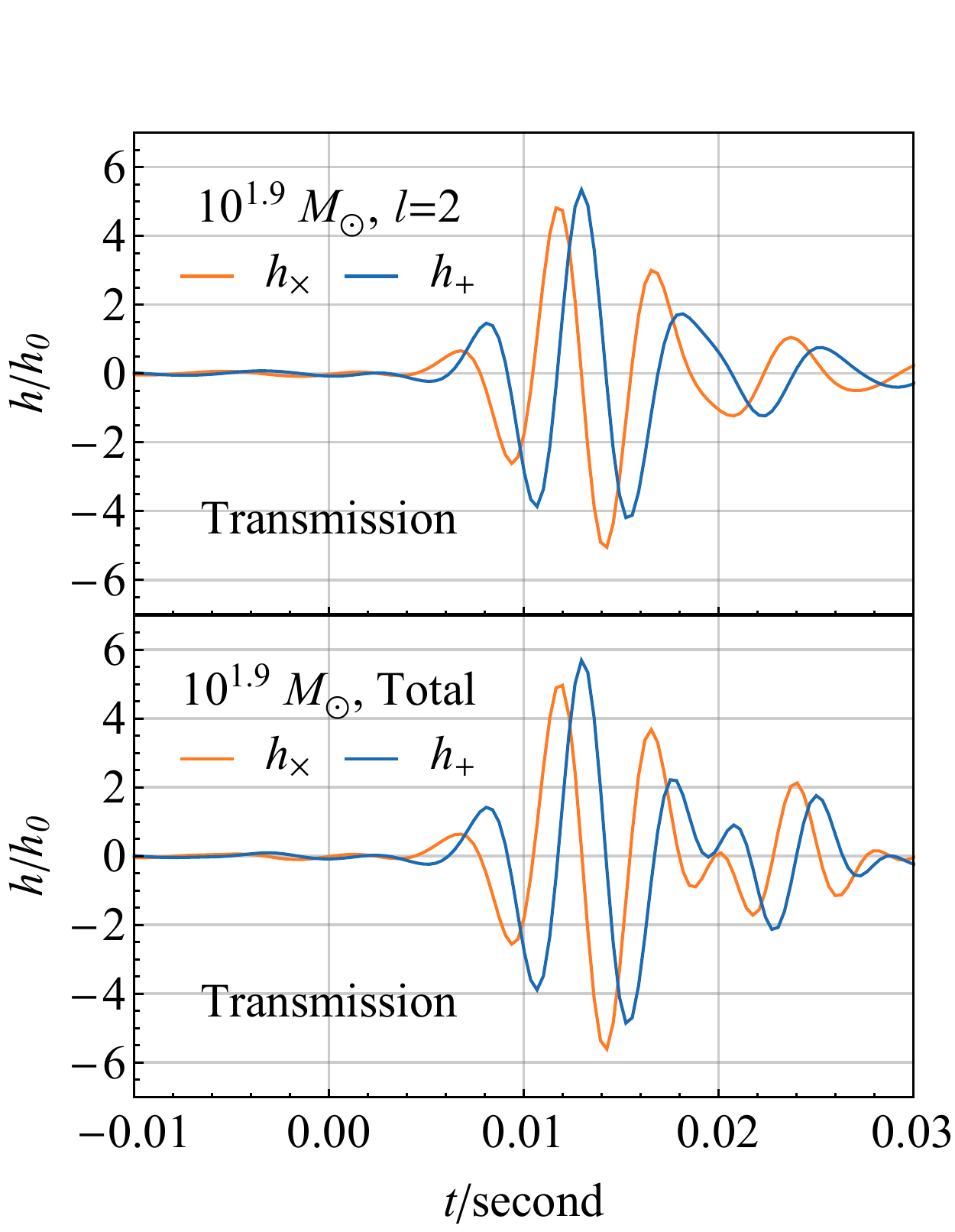} 
     \includegraphics[width=0.24\textwidth]{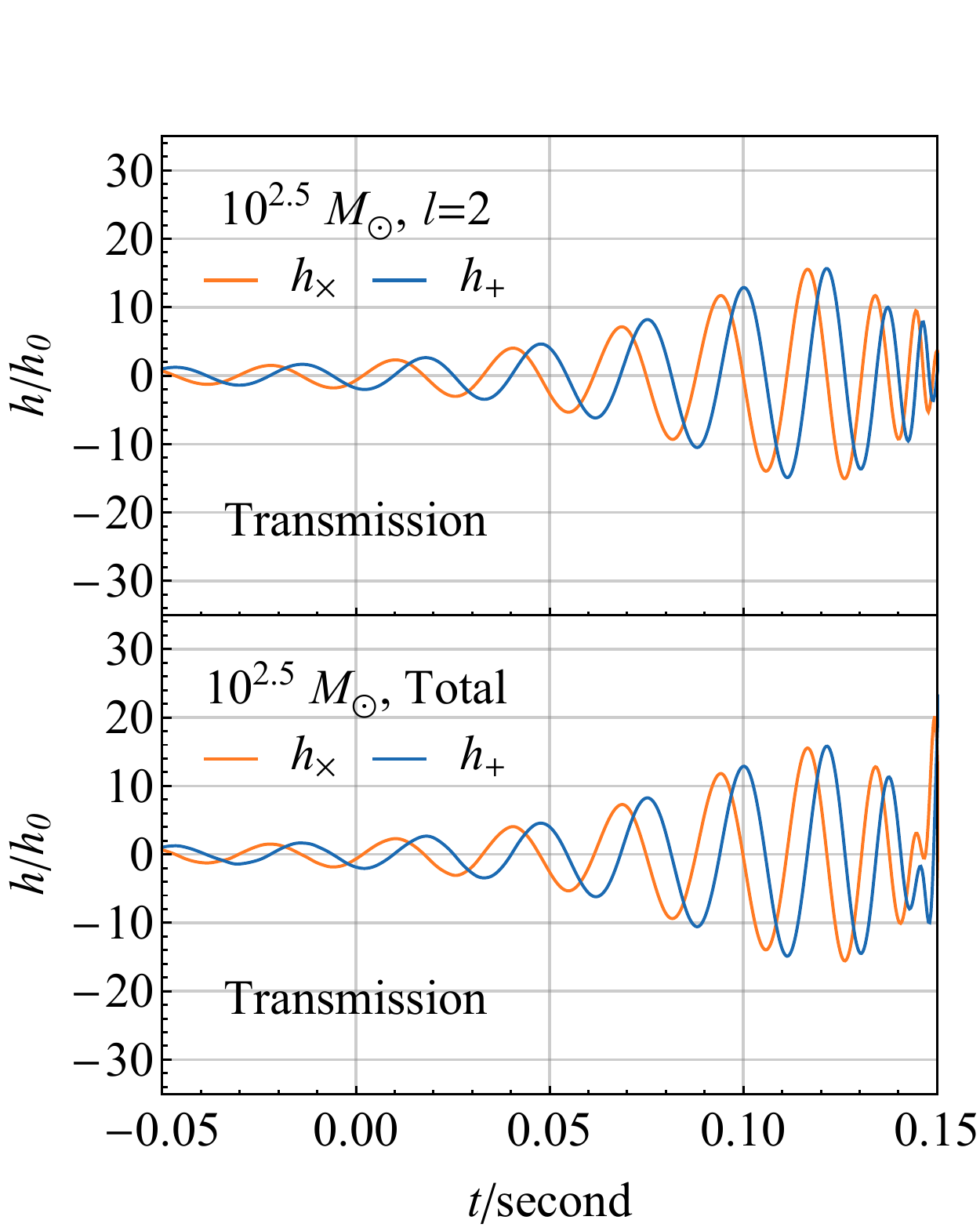}
    \caption{ Comparison of the transmitted waveform to the lowest partial wave. The wormhole masses are $10^{1.9}M_\odot$ (left) and $10^{2.5}M_\odot$ (right). Other parameters are the same as those in Fig.~\ref{fig:ref_tran_38} for comparison. For heavier wormholes, more partial waves contribute to the transmitted waveform, hence the destructive interference is more severe. Nonetheless, the first several peaks are always dominated by the lowest partial wave.}
    \label{fig:sumvsl2}
\end{figure}

Since the potentials $V^l_\text{e/o}$ scale as $1/r_g^2$, heavier wormholes have lower potentials and more partial waves transmit. Worse interference happens as a result. We find there are time ranges that only the lowest several partial waves dominate. In Fig.~\ref{fig:sumvsl2}, we compare the summed transmitted waveform to the lowest partial wave with $l=2$. The first several peaks of the transmitted wave are dominated by the lowest partial wave. This is because the resonance region of $l=2$ is the earliest to be reached by the wave train. The configuration of these early peaks can be used to extract the information of the wormhole. This conclusion is also valid for the reflected waves. Conservatively speaking, this signature serves at least as a signal that different partial waves do not arrive at the same time. To confirm it is from a wormhole, it has to be combined with other signatures, such as the isolated-chirp or anti-chirp signals discussed above.

\begin{figure}
    \centering
    \includegraphics[width=0.48\textwidth]{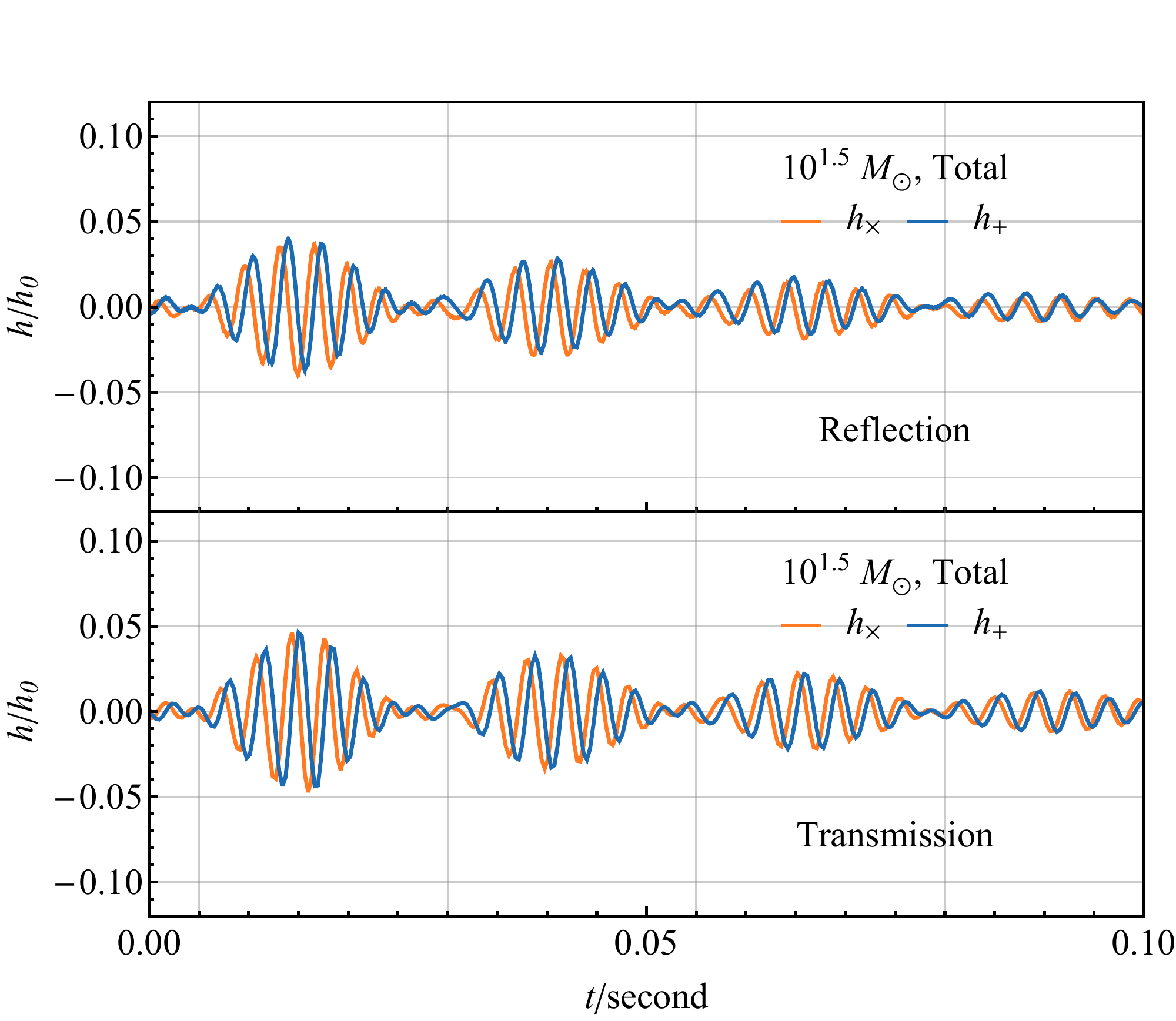}
    \caption{The echoes of reflected (top) and transmitted (bottom) waves for small-mass wormhole mass.  The wormhole mass is $M_{\wh}=10^{1.5}M_\odot$. Other parameters are the same as those in Fig.~\ref{fig:ref_tran_38} for comparison.
}
    \label{fig:echoes}
\end{figure}

Smaller-mass wormhole potentials have higher peaks and fewer partial waves contribute. One would expect the echoes are more obvious. Fig.~\ref{fig:echoes} shows the reflected and transmitted waves for wormhole mass $10^{1.5}M_\odot$. The destructive interference is less severe, which leads to more recognizable echoes. Nonetheless, the amplitudes are also smaller by an order of magnitude compared to the echoes in Fig.~\ref{fig:ref_tran_38}, which are from wormhole mass $10^{1.9}M_\odot$. In general, the echoes from wormholes with smaller masses are more challenging in observation, taking all other parameters to be the same. This conclusion also works for the reflected echoes. The incident wave is dominantly reflected by the left peak of the potentials $V^l_\text{e/o}$, with different phase shifts for different partial waves. In this sense, the best signature of a small-mass wormhole seems to be the waveform of the directly reflected wave. However, this reflected waveform is likely to be indistinguishable from that of a black hole with the same mass. To probe these small-mass wormholes, it is better to consider GW sources with higher frequencies.

Besides the wormhole mass, the potentials also depend on the parameter $\Delta r_0$, which determines the location where the two Schwarzschild metrics are sewed together. We repeat all the calculations with $\Delta r_0=10^{-10}r_g$. The waveforms are almost the same, only with the time intervals between consecutive echoes increased. 
Take $h_+$ of the reflected waveform in Fig.~\ref{fig:echoes} as an example. The first and the last peaks of the first echo appear at 0.006~second and 0.023~second, while the first dip and the last peak of the second echo appear at 0.031~second and 0.050~second. The numbers for the reflected $h_+$ waveform with $\Delta r_0=10^{-10}r_g$ are 0.015, 0.032, 0.047 and 0.066, respectively. Hence if more than one echo could be observed, this would be important to judge whether the signal is from a wormhole.

\begin{figure}
    \centering
    \includegraphics[width=0.47\textwidth]{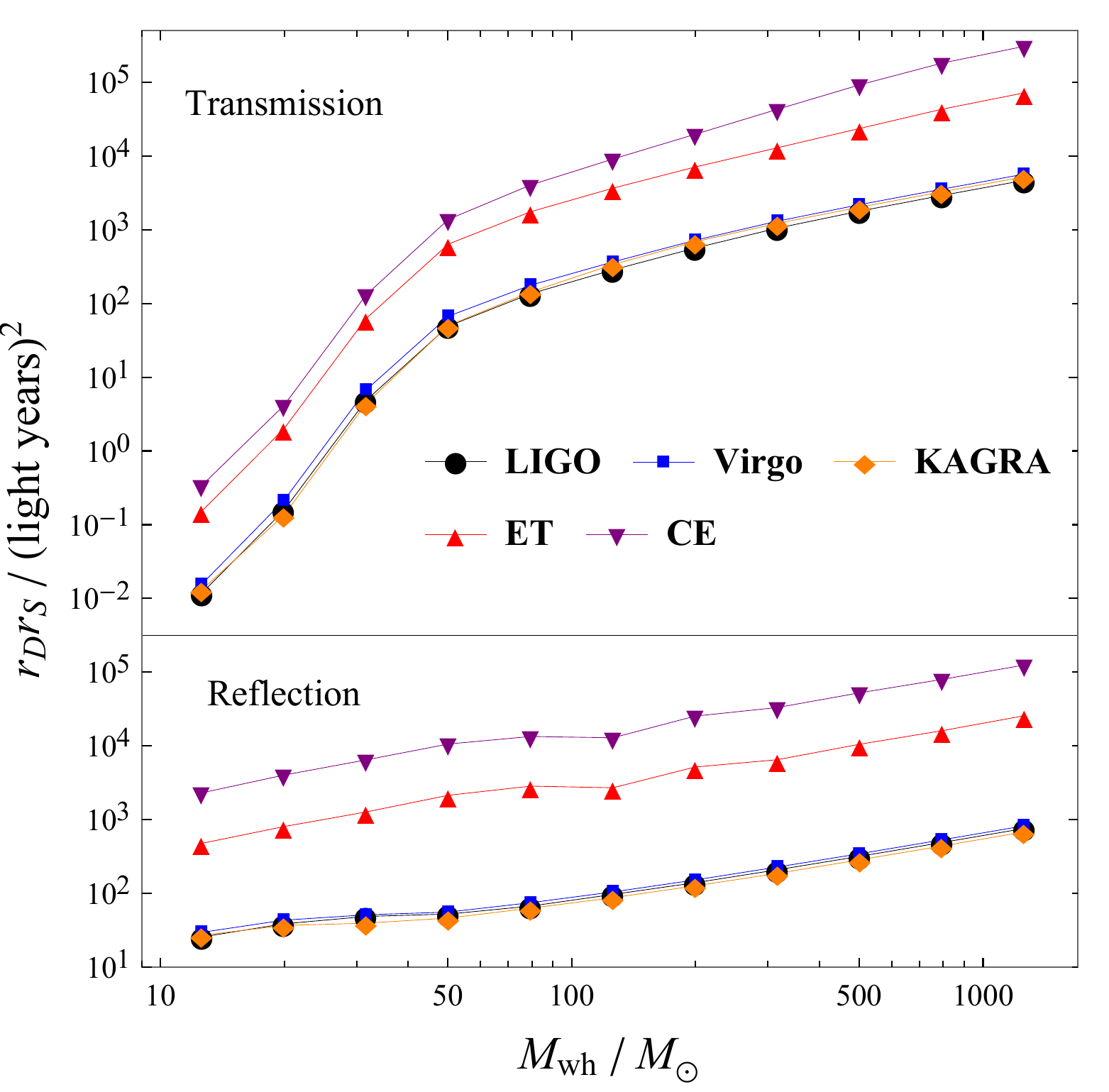} 
    \caption{
    Maximum searching ranges for the current and projected GW telescopes. The SNR is set to be 8.}
    \label{fig:range}
\end{figure}

The amplitudes of the waveforms depend inversely on the multiplication of the distance from the source to the wormhole ($r_S$) and the distance from the wormhole to the detector ($r_D$). In Fig.~\ref{fig:range}, we study the angular-averaged SNR for different wormhole masses from $10 M_\odot$ to $10^3 M_\odot$. The curves show the maximum searching ranges determined by setting $\text{SNR}=8$ for the current and projected GW telescopes, including Einstein Telescope (ET)  \cite{Punturo:2010zza} and Cosmic Explorer (CE) \cite{Evans:2016mbw}. The ranges are based on the source strength of GW150914.  With sources of similar strength, the current GW detectors can be used to explore the existence of wormholes not far from us. With a stronger and/or long-duration source, the ranges are extented accordingly. On the other hand, the more sensitive GW detectors in the future will greatly increase the searching scope.

Although this work focuses on wormholes that could be detected by ground-based interferometers, it is expected that space-borne detectors could also observe similar reflected and transmitted waveforms. In particular, the isolated-chirp, anti-chirp and recognizable echo signals can be detected by LISA \cite{Seoane:2013qna}, Taiji \cite{Taiji2017} and TianQin \cite{Luo:2015ght} if the wormhole mass is between $10^{5.5 }M_\odot$ and $10^{6.5 }M_\odot$. For BBO \cite{Corbin:2005ny} and DECIGO \cite{Seto:2001qf}, this wormhole mass range is between $10^{4.5} M_\odot$ and $10^{5.5 }M_\odot$. Thus, the combination of these GW detectors covers a very large wormhole mass range. The combined analysis of signals in different detectors would also help to confirm whether a GW event is scattered off a wormhole.


\section{Summary}\label{sec:summary}

In this work, we have studied the GW scattered off symmetric Schwarzschild wormholes. 
A time-independent scattering theory has been developed.
To our knowledge, this is the first time such a theory is applied to GW scattering.
Compared to the previous time-dependent methods, this new framework has two advantages. Firstly, the picture of partial-wave scattering is very straightforward, and the destructive interference of different partial waves is clear to see. Secondly, the calculation is simple and efficient, with the numerical error easily suppressed to a very small value. With these advantages, this new calculation framework is capable of constructing accurate templates of GWs scattered off exotic compact objects for the current and projected gravitational telescopes. 

With this new method, the scattering of different partial waves could be studied separately.
We have shown that low-frequency waves are dominantly reflected, and high-frequency waves dominantly pass through the wormhole to the other side of the throat. 
For the waves with frequencies similar to the height of a partial-wave potential, resonant scattering is observed, producing a series of echoes at the rear of both the reflected and the transmitted wave trains.

Then we have studied a real-life scattering event, using the initial GW packet generated by \verb+PyCBC+ with the masses of the black holes the same as GW150914.
Both the reflected and the transmitted waves present rich features.
Specifically, the transmitted wave consists of an isolated chirp and a series of echoes.
While the reflected wave has an anti-chirp as well as echoes.
In the future, when the GW scattering is employed to search for compact objects, these features will be useful to confirm whether it is a wormhole.

To test the feasibility of using this method to search for wormholes, we further studied the searching ranges of the current and projected GW telescopes.  With sources of similar strength as GW150914, the current GW detectors can be used to explore the existence of wormholes not very far from us. With a stronger and/or long-duration source, the ranges are extended accordingly. In the future, the more sensitive GW detectors such as LISA, Taiji, and TianQin will greatly increase the search scope. 


\appendix 

\section{Spherical partial wave expansion of a plane wave.} \label{app:plane-wave}
We adopt the polar and axial spherical harmonic matrices $M^{lm}_{s;\mu\nu}$ defined in Ref.~\cite{Berti:note},
\begin{subequations}
\begin{align}
  M^{lm}_{0;tt}&=\sqrt 2M^{lm}_{1;tr}=M^{lm}_{2;rr}=Y^{lm},\\
  M^{lm}_{3;tA}&=M^{lm}_{4;rA}=\frac{n_lr}{\sqrt 2}Y_A^{lm},\\
  M^{lm}_{5;tA}&=M^{lm}_{6;rA}=-\frac{in_lr}{\sqrt{2}}X_A^{lm},\\
  M^{lm}_{7;AB}&=\sqrt{2}m_lr^2 Y^{lm}_{AB},\\
  M^{lm}_{8;AB}&=\frac{r^2}{\sqrt 2}\gamma_{AB}Y^{lm},\\
  M^{lm}_{9;AB}&=-i\sqrt 2m_lr^2 X_{AB}^{lm},
  \end{align}
\end{subequations}
where $A,B = \theta, \phi$, and $n_l=1/\sqrt{l(l+1)}$. Other components either vanish or are related to the ones shown above by symmetry. The plane wave with definite handedness can then be expanded in partial waves,
\begin{equation}
  \label{eq-sph-exp-e}
  e^{(P)}_{\mu\nu}e^{ik r\cos \theta}=\sum_{s=0}^9\sum_{l\ge2}\sum_{m=-l}^l\mathscr C_{s;lm}^PM^{lm}_{s;\mu\nu},
\end{equation}
where $\mathscr C^P_{s;lm}$ are the expansion coefficients for parity $P$. With the orthogonality property of the spherical harmonic matrices, the non-vanishing coefficients are,
\begin{subequations}
\begin{align}
\mathscr C_{2;lm}^P &=\frac{1}{\sqrt{2}\, m_l}\mathcal F_{lm}^P\frac{j_l}{(\omega r)^2},\\
\mathscr C_{4;lm}^P &=\frac{n_l}{m_l}\mathcal F_{lm}^P\frac{(l+1)j_l-\omega rj_{l+1}}{(\omega r)^2},\\
\mathscr C_{6;lm}^P &=i^{-P}\frac{n_l}{m_l}\mathcal F_{lm}^P\frac{j_l}{\omega r},\label{eq-clm}\\
\mathscr C_{7;lm}^P & =\mathcal F_{lm}^P\frac{[l^2+3l+2-2(\omega r)^2]j_l-2\omega rj_{l+1}}{2(\omega r)^2},\label{eq-flm}\\
\mathscr C_{8;lm}^P &=-\frac{1}{\sqrt{2}}\mathscr C_{2;lm}^P,\\
\mathscr C_{9;lm}^P &=i^{-P}\mathcal F_{lm}^P\frac{(l+2)j_l-\omega r j_{l+1}}{\omega r},
\end{align}
\end{subequations}
where the argument $\omega r$ of all the spherical Bessel functions are suppressed for compactness. The gauge-invariant functions $\Psi^{lm}_\text{e/o}$ are then constructed as linear combinations of the coefficients $\mathscr C^P_{s;lm}$ \cite{Martel:2005ir},
\begin{widetext}
\begin{subequations}
\begin{align}
\Psi_\text{o}^{lm} \text{ for } e^{(P)}_{\mu\nu}e^{ik r\cos\theta}
&=
-i^P\sqrt 2m_l\mathcal F_{lm}^P\, r\, j_l(\omega r),\\
\begin{split}
\Psi_\text{e}^{lm} \text{ for } e^{(P)}_{\mu\nu}e^{ik r\cos\theta}
&=
-\frac{\sqrt 2 \, n_l^2}{m_l \Lambda}\mathcal F_{lm}^P
\bigg\{
\frac{(l^2+l+1)}{(l-1)(l+2)}\frac{r_g}{\omega r} \, j_{l+1}(\omega r)\\
&\hspace{3cm}+
\Big[(\omega r)^2 + \frac{3\, \omega^2  r_g r}{(l-1)(l+2)} - \frac{(l^2-l+3)}{l-1}\frac{r_g}{r}- \frac{2\,r_g^2}{r^2}\Big]\frac{r}{(\omega r)^2} \, j_l(\omega r)
\bigg\}.
\end{split}
\end{align}
\end{subequations}
\end{widetext}
The asymptotic behaviour of $\Psi_\text{o}^{lm}$ at $r\to +\infty$ is,
\begin{align}
-i^P\sqrt2m_l\mathcal F_{lm}^P\frac{\sin(\omega r-l\pi/2)}{\omega}.
\end{align}
The behavior for $\Psi_\text{e}^{lm}$ is the same, only with the factor $i^P$ dropped. Since in the up universe $r_*<0$, the $r$ in the results above should be replaced by $-r$.

\section{Solving for the Eigenstates.}\label{app:WKB}

Both the numerical Wronskian method and the WKB approximation are used to calculate the frequency eigenstates of the double-barrier potential scattering in Eq.~\eqref{eq-mseq-e}. The numerical method with the Wronskian determinant for double-barrier scattering is explained in detail in Ref.~\cite{Fernandez:2011}. Since the WKB approximation is much faster, it is used for most of the calculation, while the numerical method is only applied to monitor the error of the WKB method, and in the extremely small energy region where the WKB method has a relatively large error. Below we give the details of the WKB method used in this work.

The WKB approximation is further separated into two formulas, depending on the size of the frequency $\omega$ compared to the critical frequency $\omega_c$. When the frequency is small and transmission is classically forbidden, the textbook WKB formula is adopted. The WKB exponentials on both sides of a classical turning point are connected by the Airy functions. In this work, we consider the potential with two equal-height peaks, so there are in total four classical turning points. We normalize the incident wave to be $\exp(i\omega x)$ at $x\to -\infty$. In the second case when the energy is large and the transmission is allowed classically, we use a different condition to connect the WKB approximate exponentials on both sides of a peak. Consider a peak at position $x_1$ with height $V_0$. Close to the top of the peak, the potential can be Taylor expanded in $(x-x_1)$ as $V(x) = V_0 - a (x-x_1)^2$, where higher-order terms of $(x-x_1)$ are neglected. The solution is a linear combination of two parabolic cylinder functions $U\left( \pm \frac{i b}{2},(\pm 1 +i) y\right)$, where $y=a^{1/4}(x - x_1)$ and $b=a^{-1/2}(\omega^2-V_0) $. Their asymptotic behaviors are,
\begin{subequations}\label{eq:WKB-U-Asymp}
\begin{align}
    \lim_{y\to +\infty}\!\! U\left(\pm\frac{ib}{2},y_1\right) \!\! &=y_1^{-(1\pm ib)/2}  e^{- y_1^2/4},
\end{align}
\begin{align}
\begin{split}
    \lim_{y\to -\infty} U\left(\pm\frac{ib}{2},y_1\right)\!\! &=
    y_1^{-(1\pm ib)/2}  e^{-y_1^2/4}\\
    &\hspace{-2cm}
    -\frac{i\sqrt{2\pi}\, e^{\mp b\pi/2}}{\Gamma\left(\frac{1}{2}(1\pm ib)\right)}
    y_1^{-(1\mp ib)/2} e^{ y_1^2/4},
\end{split}
\end{align}
\begin{align}
\begin{split}
    \lim_{y\to +\infty} U\left(\pm\frac{ib}{2},-y_1^*\right)\!\! &=
   \left(-y_1^*\right)^{-(1\pm ib)/2}  e^{-y_1^{*2}/4}\\
    &\hspace{-2.5cm}
    +\frac{i\sqrt{2\pi}\, e^{\pm b\pi/2}}{\Gamma\left(\frac{1}{2}(1\pm ib)\right)}
    \left(-y_1^*\right)^{-(1\mp ib)/2} e^{ y_1^{*2}/4},
\end{split}
\end{align}
\begin{align}
    \lim_{y\to -\infty}\!\! U\left(\pm\frac{ib}{2},-y_1^*\right)\!\! &= \left(-y_1^*\right)^{-(1\pm ib)/2} e^{-y_1^{*2}/4},
\end{align}
\end{subequations}
where $y_1=(1+i)y$. The WKB phase integral can also be completed analytically,
\begin{align}
\begin{split}
    \Theta(x_1,x) &\equiv  
    \int_{x_1}^x p(x') \ud x'\\
    &= \frac{b}{4}\log \frac{(y+\sqrt{b+y^2})^2}{b} + \frac{y}{2}\sqrt{b+y^2},
\end{split}
\end{align}
where $p(x)$ is defined as $\sqrt{\omega^2-V_0 + a (x-x_1)^2}$. Then the WKB exponentials on both sides of the peak can be connected by the parabolic cylinder functions given above. Note that only the limits at $|y|\gg b$ of the analytic expression of $\Theta$ are needed. In this work, the two peaks have the same values of $V_0$ and $a$, but located at different positions $x_1$ and $x_2$, with $x_2<x_1$. We define the wave function as, 
\begin{align}
 \psi(x_0) &= \frac{1}{\sqrt{p(x_0)}} \exp\left[i\, \Theta(x_1,x_0)\right],
\end{align}
at $x_0\gg x_1$, and, 
\begin{align}
\begin{split}
 \psi(x_3) &= \frac{c}{\sqrt{p(x_3)}} \exp\left[i\,\Theta(x_2,x_3)\right] \\
 &\hspace{1cm}
 +  \frac{d}{\sqrt{p(x_3)}} \exp\left[-i\,\Theta(x_2,x_3)\right],
\end{split}
\end{align}
at $x_3\ll x_2$. Then the WKB method gives,
\begin{align}\label{eq:PA-final-2}
    \left(\begin{array}{c}
        c\,\mathcal{T}_2  \\
        d\,\mathcal{T}_2^{-1} 
    \end{array}\right)
    =
    \mathbb{N}(b)
    \left(\begin{array}{cc}
    \mathcal{T}_1^{-1}     & 0  \\
    0     &  \mathcal{T}_1
    \end{array}\right)
    \mathbb{N}(b)
     \left(\begin{array}{c}
        \mathcal{T}_0^{-1} \\
         0 
    \end{array}\right),
\end{align}
where the $\mathcal{T}$'s are, 
\begin{subequations}
\begin{align}
    \mathcal{T}_0 &= \exp\left[i \Theta(x_1,x_0)\right]\exp\left[-i\omega x_0\right],\\
    \mathcal{T}_1 &= \exp\left[i \Theta(x_2,x_1)\right],\\
    \mathcal{T}_2 &= \exp\left[i \Theta(x_3,x_2)\right]\exp\left[i\omega x_3\right],
\end{align}
\end{subequations}
and the matrix $\mathbb{N}(b)$ is,
\begin{align}
    \mathbb{N}(b) = 
    \left(\begin{array}{cc}
   \mathbb{N}_{11}(b)    &  \mathbb{N}_{12}(b) \\
    \mathbb{N}^*_{12}(b) & \mathbb{N}_{11}^*(b)
    \end{array}
    \right),
\end{align}
with,
\begin{subequations}
\begin{align}
\mathbb{N}_{11}(b) &=  \frac{\sqrt{2\pi}}{\Gamma\left((1-ib)/2\right)} \left(\frac{b}{2\,e}\right)^{-ib/2}e^{-b\pi /4},\\
\mathbb{N}_{12}(b) &= i e^{- b \pi /2}.
\end{align}
\end{subequations}
The reflection and transmission coefficients are $d/c$ and $1/c$, respectively. This formula approaches the Eikonal approximation analytically when $\omega^2\gg V_0$. Numerically, it connects smoothly to the textbook WKB formula for $\omega^2 < V_0$ at $\omega^2 = V_0$.

~

\begin{acknowledgements}

We thank Yue-Liang Wu for helpful comments and suggestions. HZ would also like to thank Shaun Hampton for the discussion about wormholes which initialize the idea of this work. SSB is supported by the Natural Science Foundation of Shandong Province (grant No. ZR2020MA094). SH is supported by Hubei Provincial and Municipal Double First-Class Initiative Startup Funding. HZ is supported by the Qilu Youth Scholar Funding of Shandong University.

\end{acknowledgements}

\bibliography{refs}

\end{document}